\documentclass[twocolumn]{aastex61}
\hypersetup{linkcolor=red,citecolor=cyan,filecolor=blue,urlcolor=black}

\usepackage{natbib}

\submitjournal{ApJ Letters}
\shorttitle{Probing the cold dust emission in the AB Aur disk: a dust trap in a decaying vortex?}
\shortauthors{Fuente et al.}

\begin{document}

\title{Probing the cold dust emission in the AB Aur disk: a dust trap in a decaying vortex?
\footnote{Based on observations carried out with the IRAM  Northern 
Extended Millimeter Array (NOEMA). IRAM is supported by INSU/CNRS (France), MPG (Germany), and IGN (Spain).}}

\correspondingauthor{Asunci\'on Fuente}
\email{a.fuente@oan.es}

\author[0000-0001-6317-6343]{Asunci\'on Fuente}
\affil{Observatorio Astron\'omico Nacional (OAN,IGN), Apdo 112, E-28803 
Alcal\'a de Henares, Spain}

\author{Cl\'ement Baruteau}
\affiliation{IRAP, Universit{\'e} de Toulouse, CNRS, UPS, Toulouse, France}

\author{Roberto Neri}
\affiliation{Institut de Radioastronomie Millim\'etrique (IRAM),
  300 rue de la Piscine, 38406 Saint Martin d'H\`eres, France}

\author{Andr\'es Carmona}
\affiliation{IRAP, Universit{\'e} de Toulouse, CNRS, UPS, Toulouse, France}

\author{Marcelino Ag\'undez}
\affiliation{Instituto de Ciencia de Materiales de Madrid (ICMM-CSIC), E-28049,
 Cantoblanco, Madrid, Spain}

\author{Javier R. Goicoechea}
\affiliation{Instituto de Ciencia de Materiales de Madrid (ICMM-CSIC), E-28049,
 Cantoblanco, Madrid, Spain}

\author{Rafael Bachiller}
\affiliation{Observatorio Astron\'omico Nacional (OAN,IGN), Apdo 112, E-28803 
Alcal\'a de Henares, Spain}

\author{Jos\'e Cernicharo}
\affiliation{Instituto de Ciencia de Materiales de Madrid (ICMM-CSIC), E-28049,
 Cantoblanco, Madrid, Spain}

\author{Olivier Bern\'e}
\affiliation{IRAP, Universit{\'e} de Toulouse, CNRS, UPS, Toulouse, France}

\begin{abstract}  
  One serious challenge for planet formation is the rapid inward drift
  of pebble-sized dust particles in protoplanetary disks. Dust
  trapping at local maxima in the disk gas pressure has received much
  theoretical attention but still lacks observational support.  The
  cold dust emission in the AB Aur disk forms an asymmetric ring at a
  radius of about 120 au, which is suggestive of dust trapping in a gas
  vortex. We present high spatial resolution (0".58$\times$0".78
  $\approx$ 80$\times$110 au) NOEMA observations of the 1.12 mm and
  2.22 mm dust continuum emission from the AB Aur disk.  Significant
  azimuthal variations of the flux ratio at both wavelengths indicate
  a size segregation of the large dust particles along the ring.  Our
  continuum images also show that the intensity variations along the
  ring are smaller at 2.22 mm than at 1.12 mm, contrary to what dust
  trapping models with a gas vortex have predicted. Our two-fluid
  (gas+dust) hydrodynamical simulations demonstrate that this feature
  is well explained if the gas vortex has started to decay due to
  turbulent diffusion, and dust particles are thus losing the azimuthal
  trapping on different timescales depending on their size. The
  comparison between our observations and simulations allows us to
  constrain the size distribution and the total mass of solid
  particles in the ring, which we find to be of the order of 30 Earth masses,
  enough to form future rocky planets.
\end{abstract}


\section{Introduction} 
\label{sec:intro}
The physical mechanism behind the formation of planetesimals remains
elusive. One of the main challenges is to avoid the rapid radial
drift of solid particles toward the star as they progressively
decouple from the gas. Planetesimal formation should therefore occur
on timescales shorter than the radial drift of solid particles, which
can be very fast for cm-sized pebbles \citep{Johansen2014}. One way to
halt the radial drift is to trap solid particles at a local maximum in
the disk gas pressure, which may occur for example at the outer edge
of the gap opened by a massive planet (e.g., \citealt{Lyra2009}) or at
the edge between regions with different levels of magnetohydrodynamic
turbulence \citep{Regaly2012,Flock2015}. These so-called dust traps
are promising locations for forming planetesimals and planetary cores,
and are therefore receiving a lot of theoretical attention.

Disk observations can bring valuable constraints on the dust trap
scenario for planetesimal formation. This is the case of transition
disks, which are protoplanetary disks that feature an inner region
with a large deficit in mid-IR and (sub-)mm emission, best understood
as a dust cavity, and which is surrounded by a bright emission ring in
the (sub-)mm \citep{Espaillat2014}. The emission ring of transition
disks is sometimes lopsided and multi-wavelength radio
observations suggest that these lopsided rings are caused by a
large-scale gas vortex that trap the large (typically mm-sized) solid
particles. This is the case for instance of the disks around Oph IRS
48 \citep{Marel2015} and HD 142527 \citep{Casassus2015}.

A large theoretical effort has been made in recent years to
investigate the formation and evolution of gas vortices in
protoplanetary disks, and to determine the dust's concentration,
growth and emission properties at such locations. This effort has been
made possible by the development of two-fluid hydrodynamical codes
that model the evolution of both the gas and dust of protoplanetary
disks. In particular, hydrodynamical simulations show that (i)
  the gaseous gap carved by a giant planet could explain the large
  dust cavities seen in the radio emission of some transition disks
  \citep{Zhu2012}, (ii) the outer edge of this gap might become
  unstable against the Rossby-wave instability (a linear instability
  setting at a minimum in the gas potential vorticity or, in practice,
  at a pressure maximum) and form a large crescent-shaped vortex
  \citep{Zhu2016}, and (iii) such large-scale gas vortices impart
  different trapping locations for small and large solid particles due
  to the gas self-gravity, with small dust grains trapped at the
  vortex's center but the larger ($>$ a few mm) particles trapped
  largely ahead of the vortex in the azimuthal direction
  \citep{Baruteau2016}. This size segregation of particles trapped in
vortices leaves imprints to the dust's continuum emission in the
(sub-)mm \citep{Baruteau2016} and may have important implications for
planetesimal formation.

The comparison between theoretical models and observations is, of
course, not straightforward. Part of the reason is that, even in
  the era of large (sub-)mm telescopes (NOEMA, ALMA), spatially
  resolving disks at sufficient sensitivity is challenging.  But also,
  interpreting the emission is uncertain due to assumptions for the
  dust temperature and opacity. Because of its large cavity and large
  (sub)mm flux, AB Aur is one of the few nearby transition disks that
  allows a fair comparison with models.

AB Aur is a nearby Herbig Ae star (A0-A1, d=145~pc,
\citealp{Hernandez2004, Gaia2016}) that hosts a well-known lopsided
transition disk. Plateau de Bure Interferometer (PdBI) images of the
CO 2$\rightarrow$1 and $^{13}$CO 2$\rightarrow$1 lines and of the
continuum emission at 1.3 mm indicate that the dust cavity
  radius in the AB Aur disk extends up to $\sim$70$-$100 au
($\sim$0".5$-$0".7; \citealp{Pietu2005,Pacheco2016}). The higher
angular resolution observations by \citet{Tang2012} showed the
existence of a compact inner disk inside the dust cavity with a small
(few degrees) inclination relative to the dust ring. The 1.3 mm
continuum emission from the ring outside the cavity is highly
non-axisymmetric suggestive of a dust trap. Spirals detected in both
the CO gas emission \citep{Tang2012,Tang2017} and in near infrared
scattered light emission \citep{Hashimoto2011} are suggestive of the
presence of at least one massive companion in the disk
\citep{Dong2016}.

We present in Sect.~\ref{sec:obs} high spatial resolution (0".58
$\times$ 0".78) NOEMA observations of the 1.12 and 2.22 mm dust
continuum emission in the AB Aur disk.  Synthetic maps of the dust
continuum emission computed from two-fluid (gas+dust) hydrodynamical
simulations are then presented in Sect.~\ref{sec:hyd}. Concluding
remarks follow in Sect.~\ref{sec:conclusions}.

\begin{figure*}
\centering
  \resizebox{0.99\hsize}{!}
  {
    \includegraphics{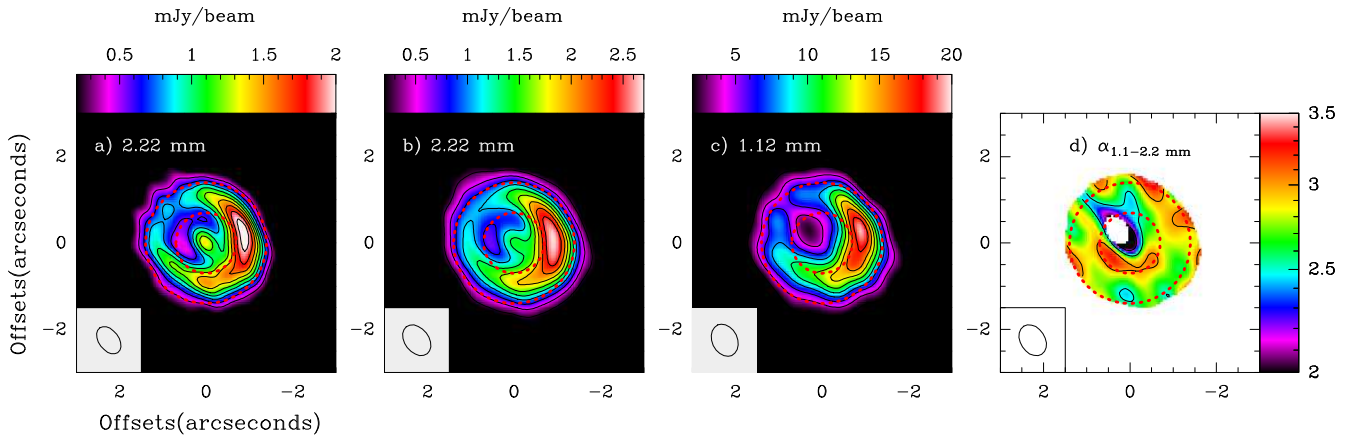}    
  }
  \caption{Continuum images at 2.22 mm and 1.12 mm with NOEMA (panels
    a, b and c). Offsets are relative to 
    RA: 04$^{\rm h}$55$^{\rm m}$45$^{\rm s}$.853, 
     Dec: 30$^\circ$33$'$03$''$.82.
    The image in panel b has been derived from the 2.22
    mm observations by applying the tapering required in the uv-plane
    to get the same spatial resolution as in the 1.12 mm image. Panel
    d shows the 1.12 mm/2.22 mm spectral index derived from panels c
    and b. Contour levels are: a) 0.38 ($\approx$8$\sigma$) to 1.98 by 0.2 mJy/beam;
      b) 0.27  ($\approx$7$\sigma$) to 2.4 by 0.27 mJy/beam;
      c) 1.95  ($\approx$6$\sigma$) to 17.56 by 1.95 mJy/beam; and
      d) 2 to 3.5 by 0.5. The uncertainty in the spectral index is of $\pm$0.29.
      The two red dashed circles have radii of
      0".7 and 1".4 and are drawn to indicate the radial extension of the ring.
      }
      
 \label{Fig1}
\end{figure*}
\section{Data reduction and observational results} 
\label{sec:obs}
Our AB Aur observations were carried out in the C-config (baselines=72-240 m)
at 266 GHz (1.12 mm) in 2016 December 4-6, and in the A-config 
(baselines=72-760 m) at 135 GHz (2.22 mm) in 2017 January 1-3. Data 
calibration and imaging were done using the package \texttt{GILDAS}\footnote{See
  \texttt{http://www.iram.fr/IRAMFR/GILDAS} for more information about
  the GILDAS softwares~\citep{Pietu2005}.}\texttt{/CLASS} software.
We tried to self-calibrate the images to improve the S/N ratio. 
However, the improvement was not significant and after some preliminary tests, we 
decided to keep the non-self calibrated images at both wavelengths, in order to avoid 
any bias in the cleaning process because of the assumed source model.
In order to optimize the spatial resolution we applied uniform
weighting to both images.  The achieved synthesized beams and
rms noise levels are 0".71$\times$0".46 PA 36$^\circ$ and 45 $\mu$Jy/beam
at 2.22 mm, and 0".78$\times$0".58 PA 33$^\circ$ and 340 $\mu$Jy/beam
at 1.12 mm.  The final images are shown in Fig.~\ref{Fig1}a
and~\ref{Fig1}c. Flux calibration uncertainty is $\sim$10\% in
  both images.

Our 2.22 mm image (Fig.~\ref{Fig1}a) shows a compact emission peak at
RA: 04$^{\rm h}$55$^{\rm m}$45$^{\rm s}$.853, Dec:
30$^\circ$33$'$03$''$.82, which agrees with the optical position of
the star and the 7~mm continuum source detected by
\citet{Rodriguez2014}. This point source is not clear in the 1.12 mm
image (Fig.~\ref{Fig1}c) because of the lower spatial resolution at
this frequency. To investigate the nature of the emission of this
compact source, we derived the cm-mm spectral index. The best
fit is obtained with a spectral index of $+$0.73$\pm$0.08, which is
consistent with the spectral indices expected for a radial thermal
(free-free) jet \citep[e.g.,][]{Reynolds1986}. Yet, some fraction of
the flux could come from a small dusty disk inside the cavity.

A lopsided ring is observed in the 1.12 and 2.22 mm continuum emission
maps. To make a reliable comparison between both emission maps,
  we produced a 2.22 mm image with the tapering required in the
  uv-plane to get the same spatial resolution as in the 1.12 mm
  image.The resulting image is shown in Fig.~\ref{Fig1}b. A
particularly interesting result is that the azimuthal contrast ratio
along the ring is smaller at 2.22 mm than at 1.12 mm (this ratio
  is $\sim$2.5$\pm$0.2 and $\sim$3.6$\pm$0.3 at 2.22 mm and 1.12 mm,
  respectively, at the same angular resolution of
  0".78$\times$0".58). This is the opposite to what is usually
expected from models of dust trapping in a gas vortex, which predict
that the larger the particles, the stronger their concentration in the
vortex and therefore the larger their emission if optically thin
\citep[e.g.][]{Pinilla2015}.  We used these images to derive the spectral 
index ($\alpha$) map defined as $F_{\nu} \propto \nu^{\alpha}$ 
(see Fig.~\ref{Fig1}d).
Only $>$10$\times \sigma$ fluxes were taken into account to compute
the flux ratio map, hence we expect that the uncertainty in our ratio
is by $\approx20\%$, which translates in an error of
    $\pm0.29$ in the 1.12 mm/2.22 mm spectral index. We detect a
  slight asymmetry in such map, which is related to the different
  azimuthal contrast ratios in the 1.12 mm and 2.22 mm images.
  
The asymmetry in the spectral index can be related with a change in the dust properties
if the emission is optically thin. At 1mm, the beam-averaged peak brightness temperature is $\sim$0.9 K. 
Assuming that the width of the dusty ring is  0".1, the real brightness temperature would be of $\sim$7.2 K, well below the dust temperature,
T$_d$=30 K \citep{Pacheco2015}, and consistent with the assumption of optically thin emission. Of course we might have some inhomogeneities within the ring where the optical depth could be larger. In our model (see
Sect. 3.2) we predict a maximum optical depth of $\sim$0.9 in the dust trap, that is barely optically thick.

\section{Hydrodynamical simulations and synthetic maps of dust emission} 
\label{sec:hyd}

\subsection{Methodology}
We carried out a two-fluid (gas+dust) hydrodynamical simulation of the
AB Aur disk using the code Dusty FARGO-ADSG, an extended version of
the public code \href{http://fargo.in2p3.fr/-FARGO-ADSG-}{FARGO-ADSG}
\citep{Masset2000, Baruteau2008} which includes dust
\citep{Baruteau2016}. We assume that the dust cavity is carved by a
2-Jupiter mass planet on a fixed circular orbit at 96 au from
the star (its mass is set to $2.4M_{\odot}$ based on the spectral
  type of AB Aur, \citealp{DeWarf2003}).  
The gas continuity and momentum equations are solved on a polar grid
with cylindrical coordinates $\{r, \varphi\}$. We use 400 grid cells
logarithmically spaced in radius between 28 and 268 au, and 600
cells evenly spaced in azimuth between 0 and 2$\pi$. A locally
isothermal equation of state is used with the gas temperature fixed in
time, given by $T(r) \approx 65\,{\rm K} \times (r/96\,{\rm
  au})^{-0.7}$. It corresponds to a disk's aspect ratio ($H/r$) of 0.1
at 96 au. Gas self-gravity is included and the effects of
turbulence are modeled by a constant alpha turbulent viscosity with
$\alpha=5\times10^{-4}$. The initial gas surface density is $\approx
1.1\,{\rm g\,cm}^{-2} \times (r/96\,{\rm au})^{-1}$ which corresponds to
a total gas mass of $\approx 1.7\times10^{-2} M_{\odot}$ \citep{Tang2012}. To avoid
reflections of the planet wakes near the grid's radial edges,
so-called wave-killing zones are used where gas fields are damped
toward their initial radial profile.

Dust is modeled as Lagrangian test particles that feel the gravity of
the star, planet and gas (since gas self-gravity is accounted for) and
gas drag. Turbulence is modeled by applying stochastic kicks to the
particles radius and azimuth at each timestep of the simulation,
following the approach of \citet{Charnoz2011} and prescribing the
dust's turbulent diffusion coefficient as in \citet{Youdin2007}.  
For computational reasons, in the simulation 
we use $10^5$ particles with a size
distribution $n_{\rm simu}(s) \propto s^{-1}$ for particle sizes $s$
between 10 $\mu$m and 10 cm. Note that the assumed grain 
size distribution does not affect the particles dynamics for a given size.
Dust self-gravity, growth, fragmentation and drag onto the gas are not
taken into account.
The particles
internal density is 2 g cm$^{-3}$, and their temperature is
that of the gas at their location.

From the spatial distribution of the solid particles in our simulation
we compute synthetic maps of the dust's continuum emission at 1.12 and
2.22 mm. This requires specifying the dust's size distribution
  $n(s)$ in the modeled disk as well as the total mass of dust. We
  take $n(s) \propto s^{-3.5}$, a minimum particle size of 10 $\mu$m
  (as in the simulation) but the maximum particle size is set as a
  free parameter (with an upper limit at 10 cm since it is the maximum
  particle size in the simulation). The total dust mass is also taken
  as a free parameter. Opacities are computed with Mie theory
assuming 60\% astrosilicates and 40\% water ices. Optical constants
for water ices are obtained from the Jena
database\footnote{\href{http://www.astro.uni-jena.de/Laboratory/Database/databases.html}{{www.astro.uni-jena.de/Laboratory/Database/databases.html}}},
those of astrosilicates are from \citet{DraineLee84}. The disk is
taken to be 145 pc away, with an inclination of 26$^\circ$ and a
  position angle of -30$^\circ$ \citep{Tang2012}.

\begin{figure*}
\centering
\gridline{
	\fig{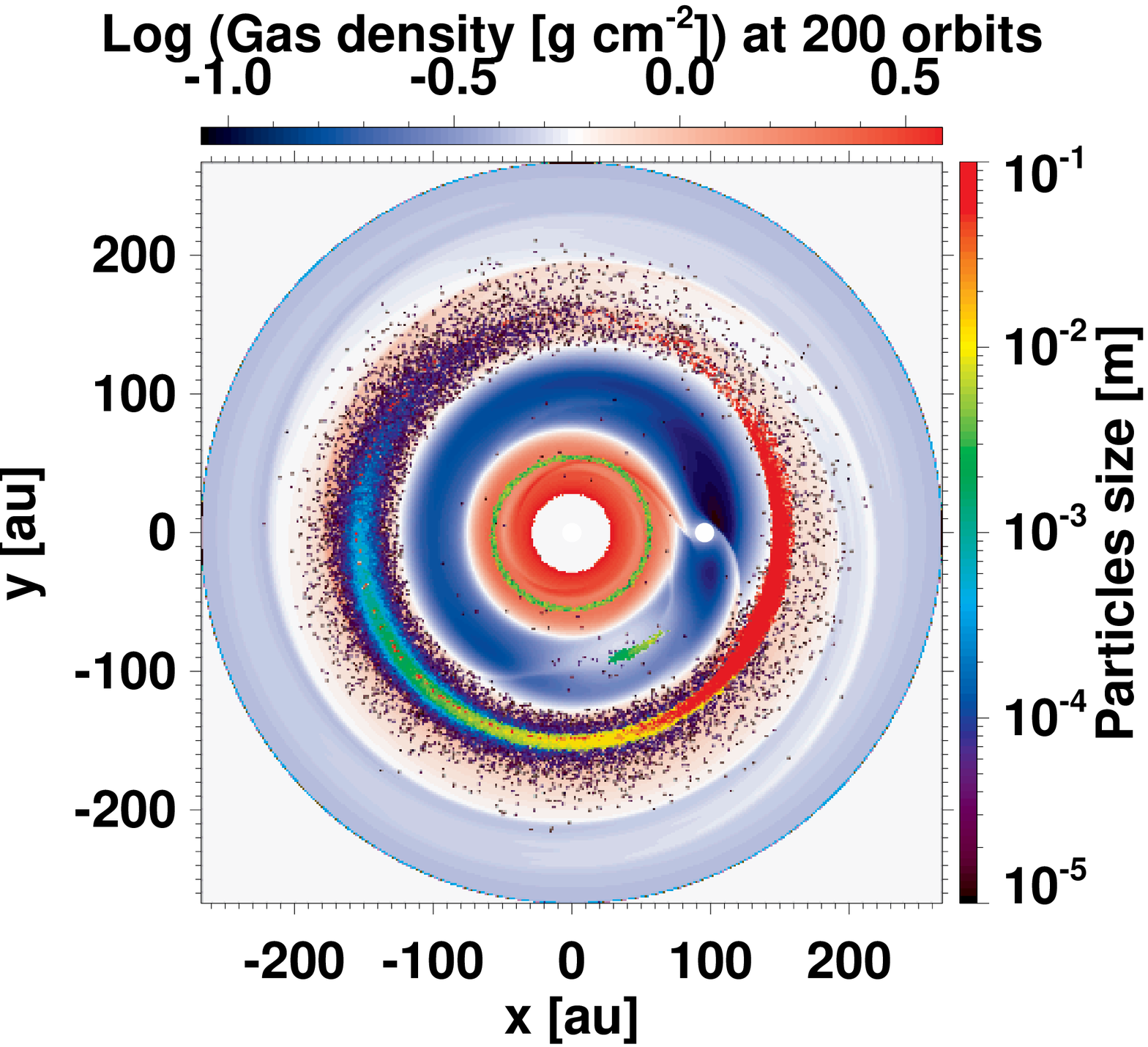}{0.27\textwidth}{(a)}
	\fig{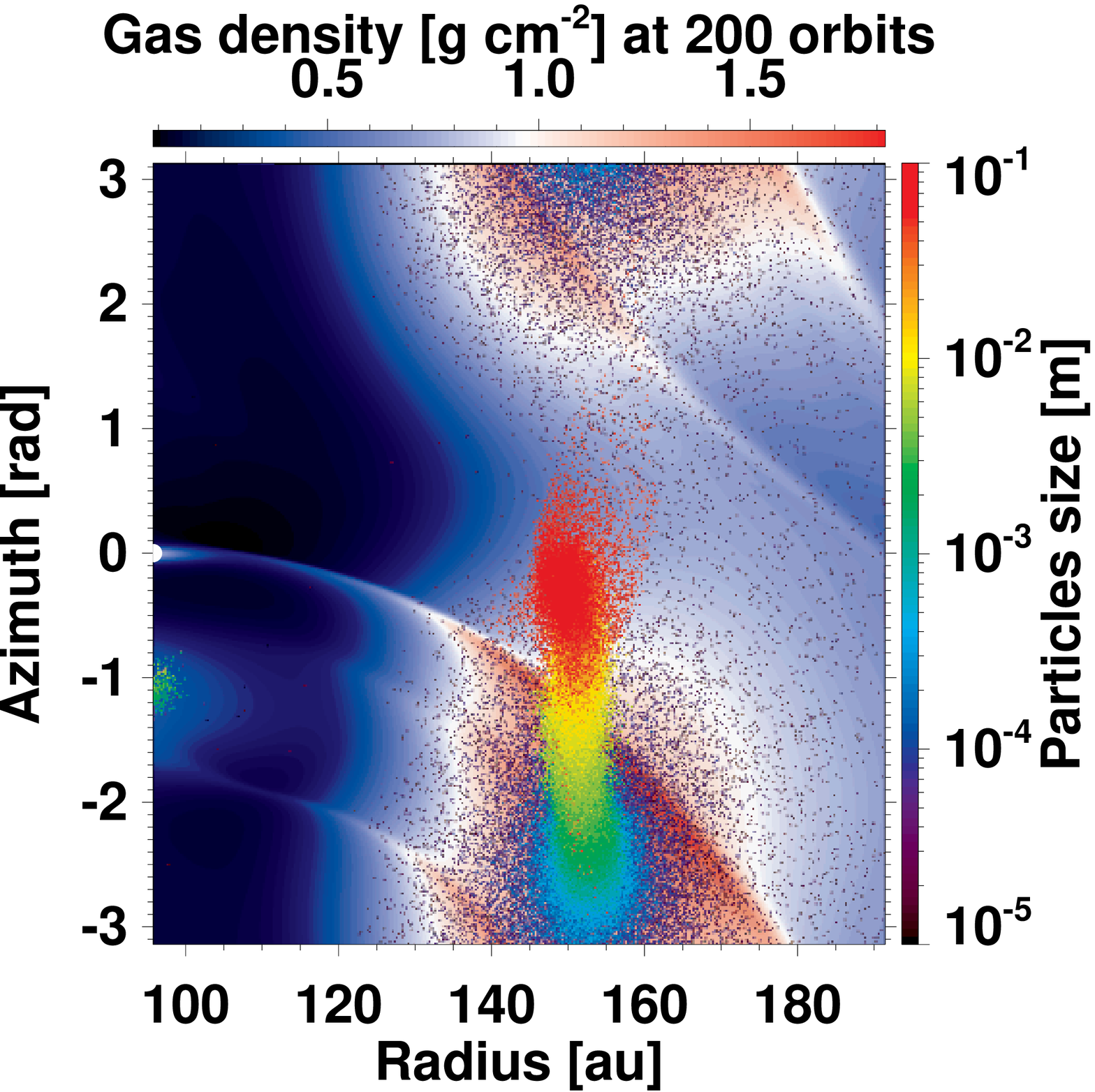}{0.24\textwidth}{(b)}
	\fig{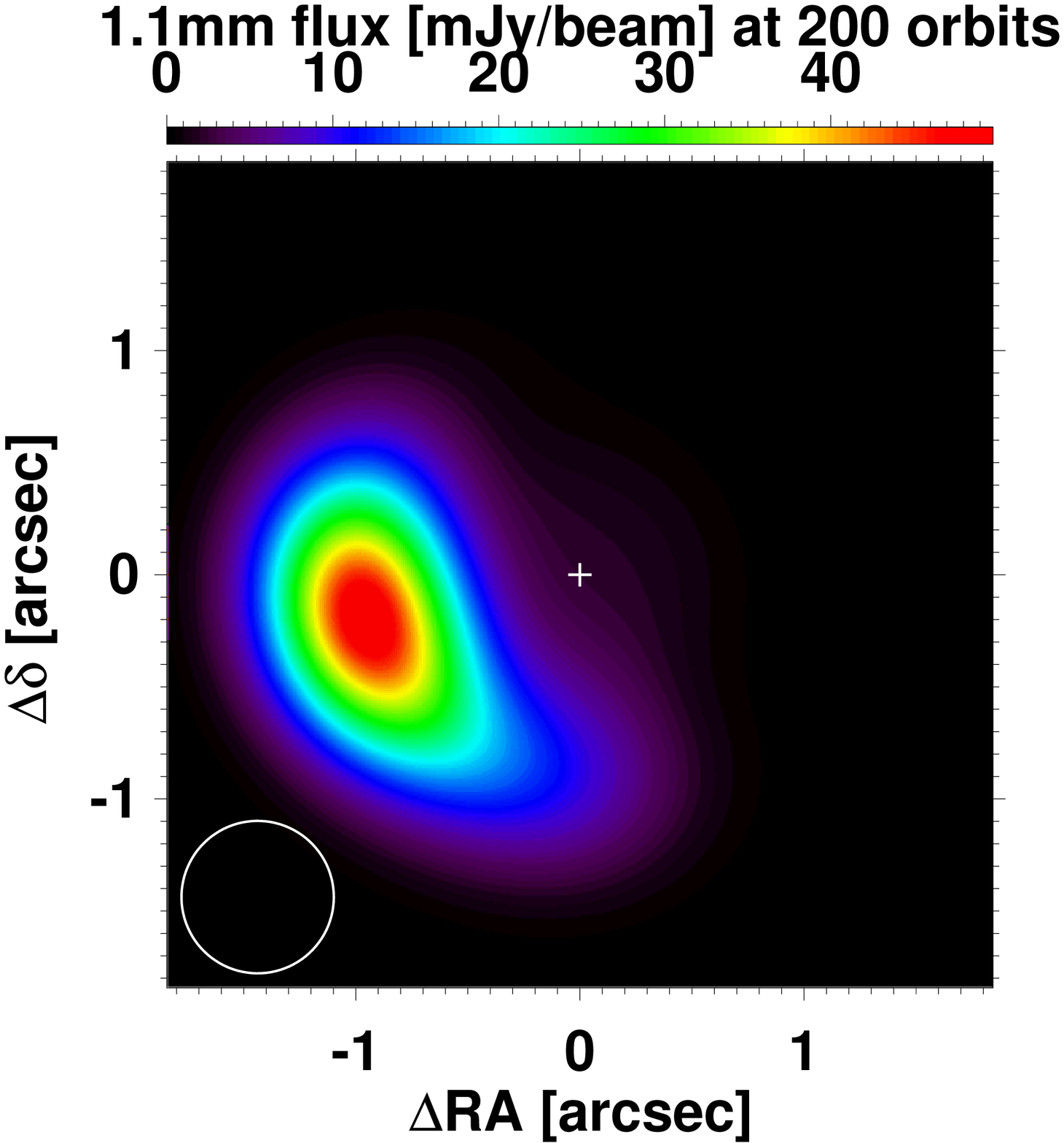}{0.24\textwidth}{(c)}
	\fig{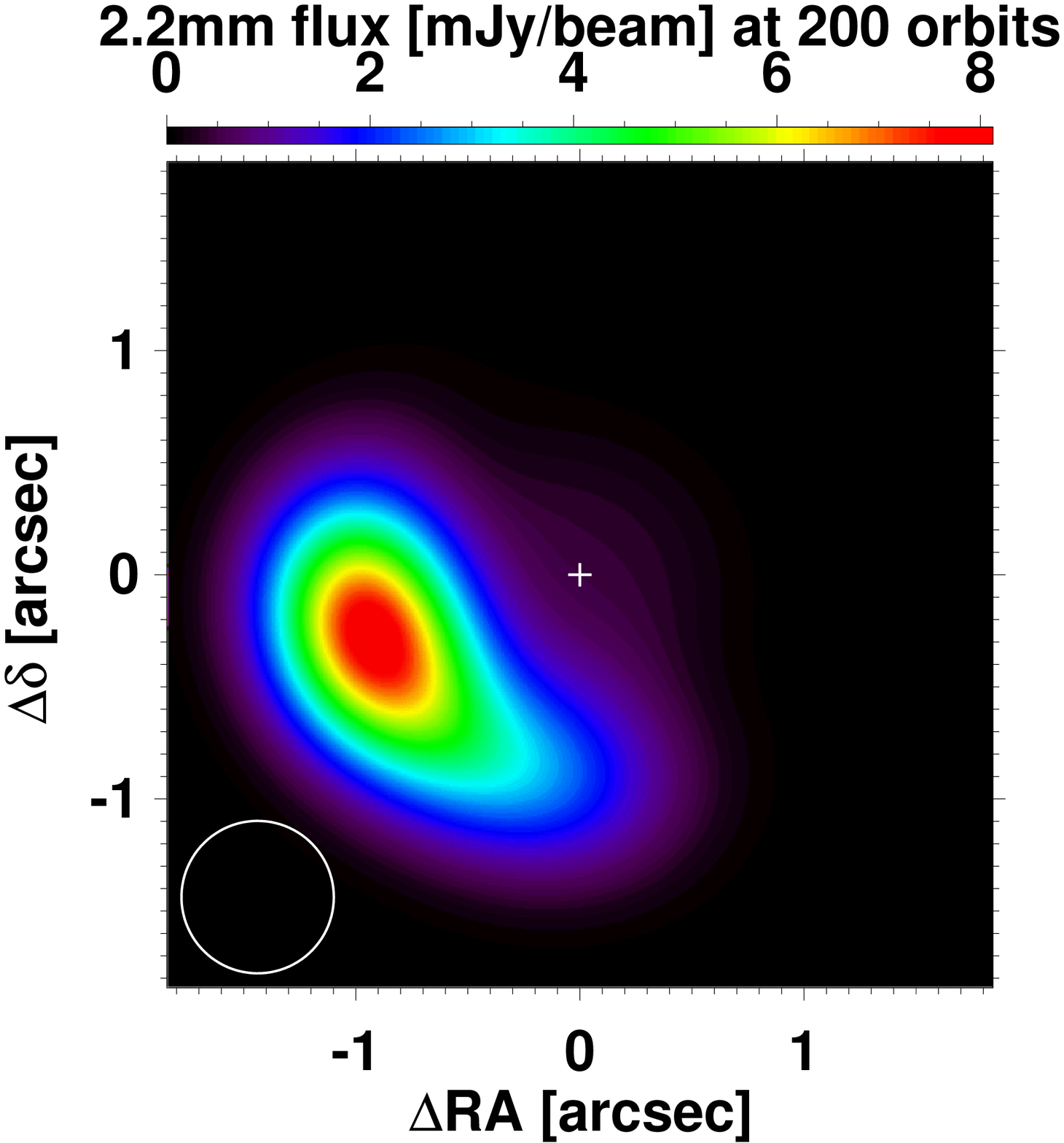}{0.24\textwidth}{(d)}
}
\gridline{
	\fig{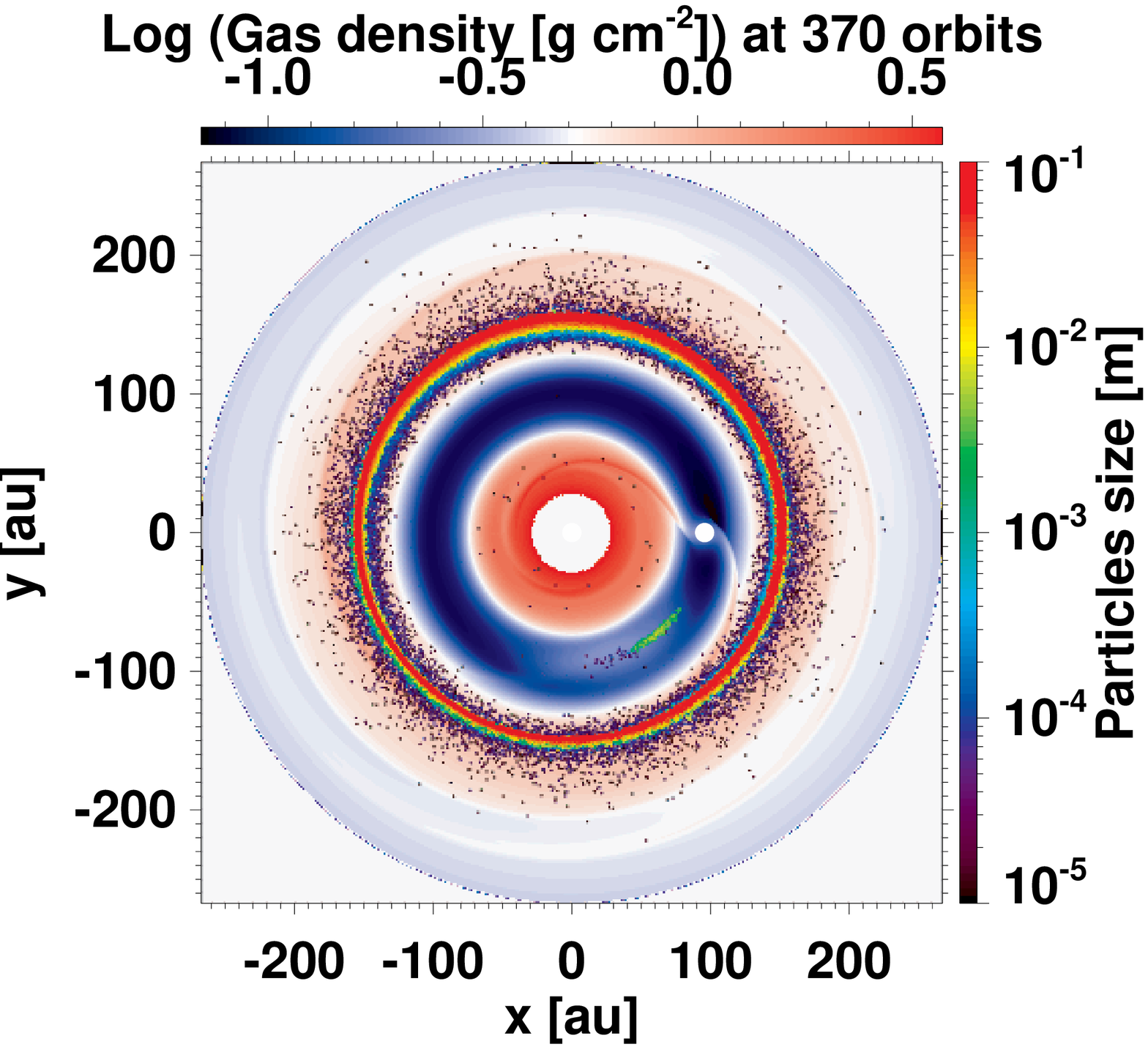}{0.27\textwidth}{(e)}
	\fig{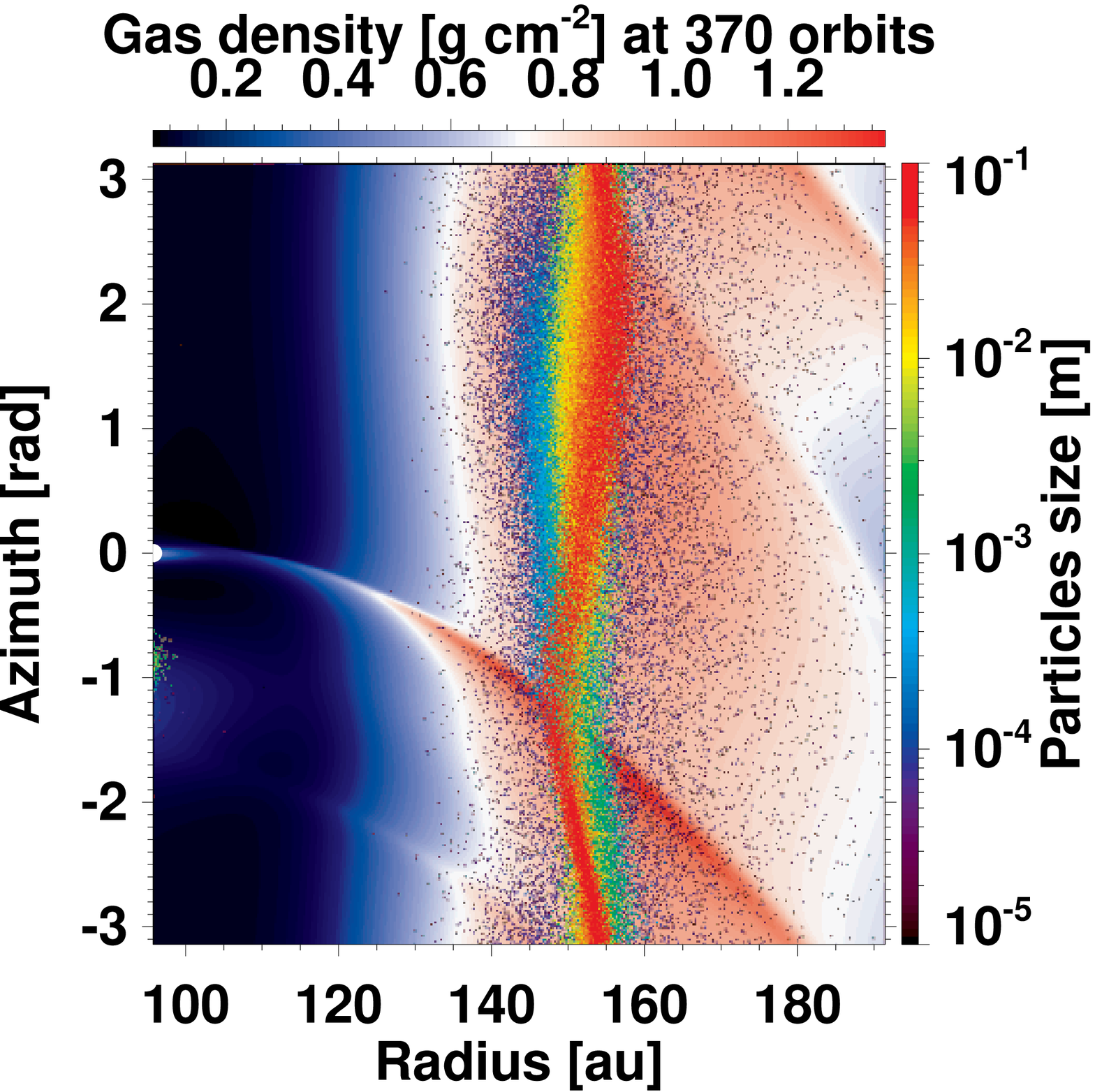}{0.24\textwidth}{(f)}
	\fig{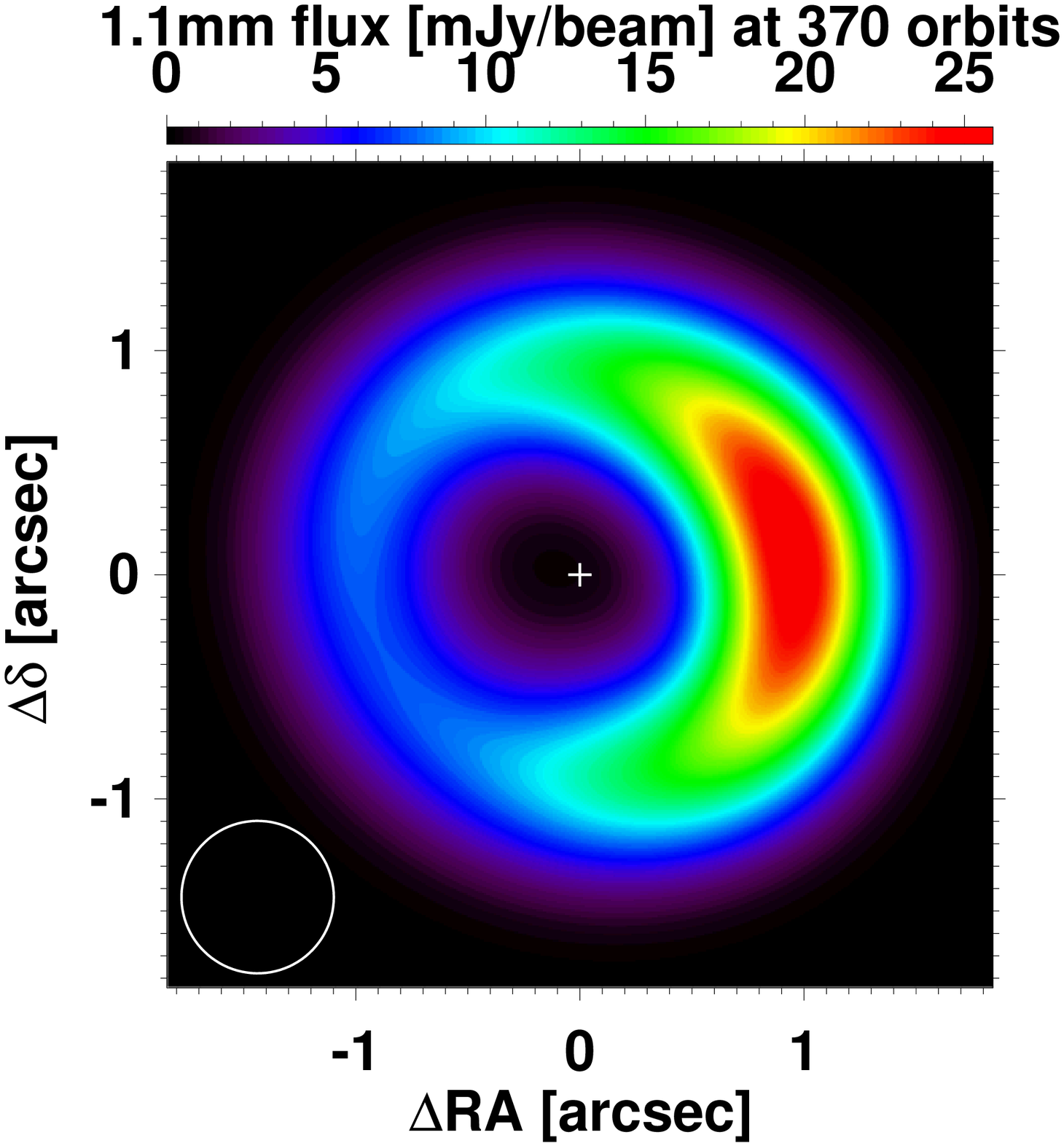}{0.24\textwidth}{(g)}
	\fig{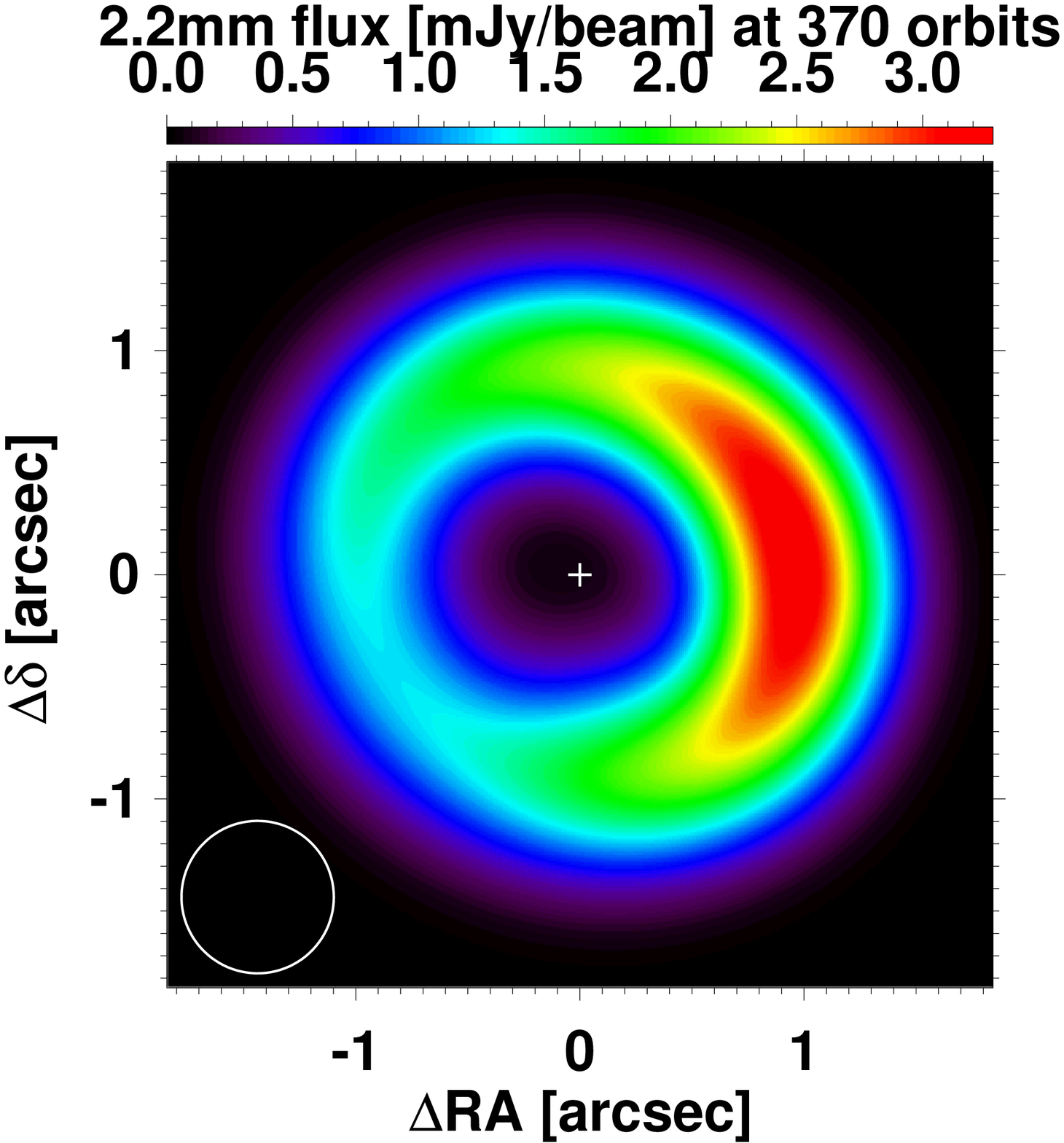}{0.24\textwidth}{(h)}
}
\caption{Results of hydrodynamical simulation modeling the AB Aur
  disk. A putative 2-Jupiter-mass planet located at 96 au from
  the central $2.4M_\odot$ star is assumed to carve the dust cavity in
  the disk.  First column: gas surface density with the location of
  solid particles shown by colored dots. The white circle marks the
  planet's location. Second column: same results displayed in polar
  coordinates with a radial zoom on the vortex location. Third and
  fourth columns: synthetic maps of the dust's continuum emission at
  1.12 mm and 2.22 mm inferred from the simulation for a dust size
  distribution $n(s) \propto s^{-3.5}$ for $s$ between 10$\mu$m and 1
  cm, and a total mass of solid particles $\approx 30M_{\oplus}$ along
  the ring. The disk is assumed to be 145~pc away with an
    inclination of 26$^\circ$ and a position angle of -30$^\circ$
    \citep{Tang2012}. The plus sign shows the star.  Flux maps are
  convolved by a Gaussian beam of 0".68 FWHM, shown by the empty circle
  in the bottom-left corner. Results are shown at two different times
  in the simulation: at 200 planet orbits (0.12 Myr), when the
  vortex's strength is roughly at its maximum level (upper panels) and
  at 370 orbits (0.22 Myr), when the vortex has started to
  decay (lower panels).
    \label{Fig2}
  }
\end{figure*}

\subsection{Results}
The planet forms a large-scale vortex in the gas due to the
Rossby-wave instability \citep{lovelace99} developing at the pressure
maximum located at the outer edge of the planet's gap in the gas (at
about 150 au). Dust particles are initially introduced between
144 and 182 au so that the large majority of them drift toward
the pressure maximum. This is meant to maximize the particles
resolution at that location, where we aim at predicting the dust's
continuum emission. Contours of the gas surface density with the
  particles location overplotted by colored dots are displayed in
  Figs.~\ref{Fig2}a and~\ref{Fig2}b after 200 planet orbits 
  (that is, at 0.12 Myr). They show that the
larger the particles, the further they are trapped ahead of the
vortex's center in the azimuthal direction. This size segregation
arises from the vortex's self-gravity \citep{Baruteau2016}.
Figs.~\ref{Fig2}c and~\ref{Fig2}d show the synthetic maps at 1.12 and
2.22 mm at the same time. They are obtained for a maximum particle size of 1 cm, and a
total dust mass of about $30M_{\oplus}$. Comparison with Fig.~\ref{Fig2}a shows 
that the synthetic emission maps peak near the vortex's center and are much less 
extended azimuthally than in the observations (compare with Figs.~\ref{Fig1}a
to~\ref{Fig1}c).

The gas vortex however has a finite lifetime and ultimately decays due
to gas turbulent diffusion. It implies that dust particles
progressively lose the azimuthal trapping effect of the vortex.  As
illustrated at 370 orbits (0.22 Myr) in Figs.~\ref{Fig2}e
and~\ref{Fig2}f, particles progressively acquire a near uniform
spatial distribution in azimuth along the pressure maximum. Particles
lose memory of their azimuthal trapping within different timescales
depending on their size, due to the combined action of
(size-dependent) gas drag and dust's turbulent diffusion (a more detailed
description of this result will be presented by \citealp{Baruteau2018}). The
synthetic maps in Figs.~\ref{Fig2}g and~\ref{Fig2}h show that, during
the vortex decay, the continuum emission forms a lopsided ring with
intensity variations along the ring that are consistent with our NOEMA
observations. The optical depth at the flux peak prior to convolution 
is $\approx0.9$ and 0.4 at 1.12 and 2.22 mm, respectively, so the 
dust's continuum emission is marginally optically thick at these positions. Comparison with 
Figs.~\ref{Fig2}c and~\ref{Fig2}d  highlights that the ring's observed 
properties can only be reproduced when the vortex has started to decay.
Note that this result is an evolutionary trend that does not depend on 
the exact values of the dust mass and grain size distribution and
suggests the presence of a decaying vortex in AB Auriga.

We point out that particles are also trapped in the Lagrangian point
located $\sim$60$^\circ$ beyond the planet's location. Their continuum
emission is a small fraction (a few percent) of the averaged flux in
the ring, and an angular resolution $\lesssim0".2$ would be required
to unambiguously disentangle the emission in the Lagrangian point from
that in the ring, which our model predicts to have a radial width of about 
15 au (0".1).

\begin{figure*}
\centering
\gridline{
	\fig{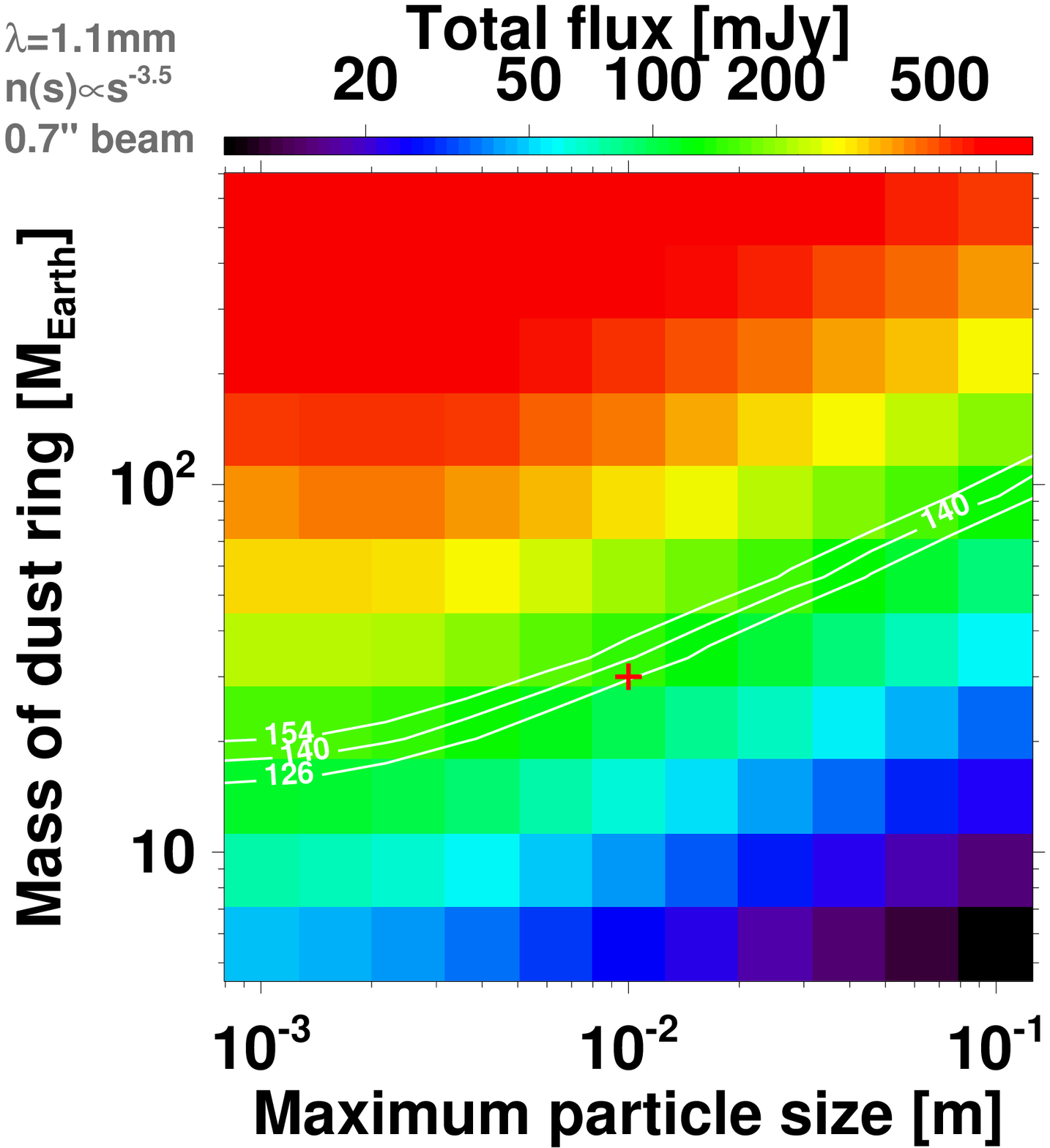}{0.45\textwidth}{(a)}
	\fig{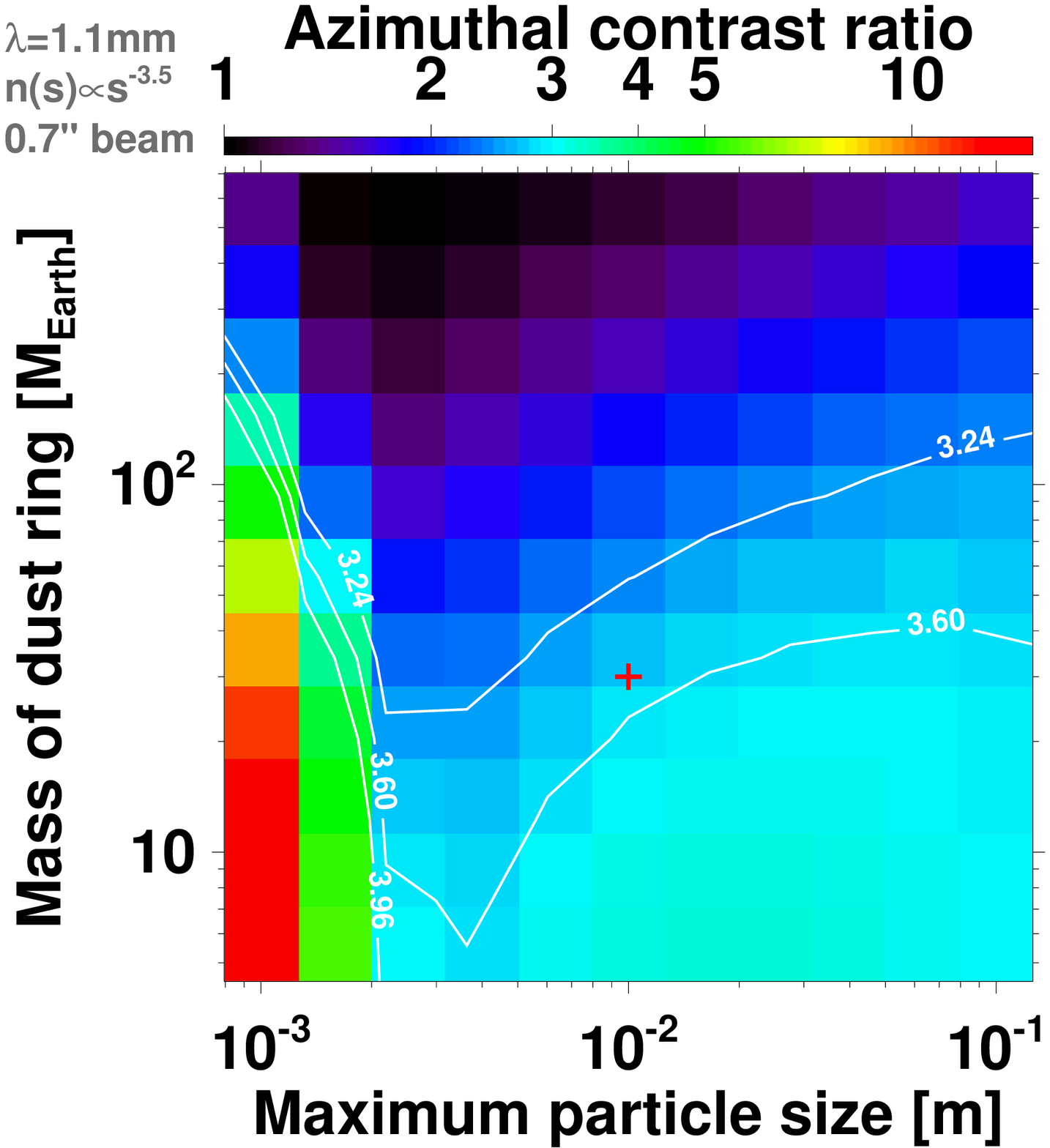}{0.45\textwidth}{(b)}
}
\gridline{
	\fig{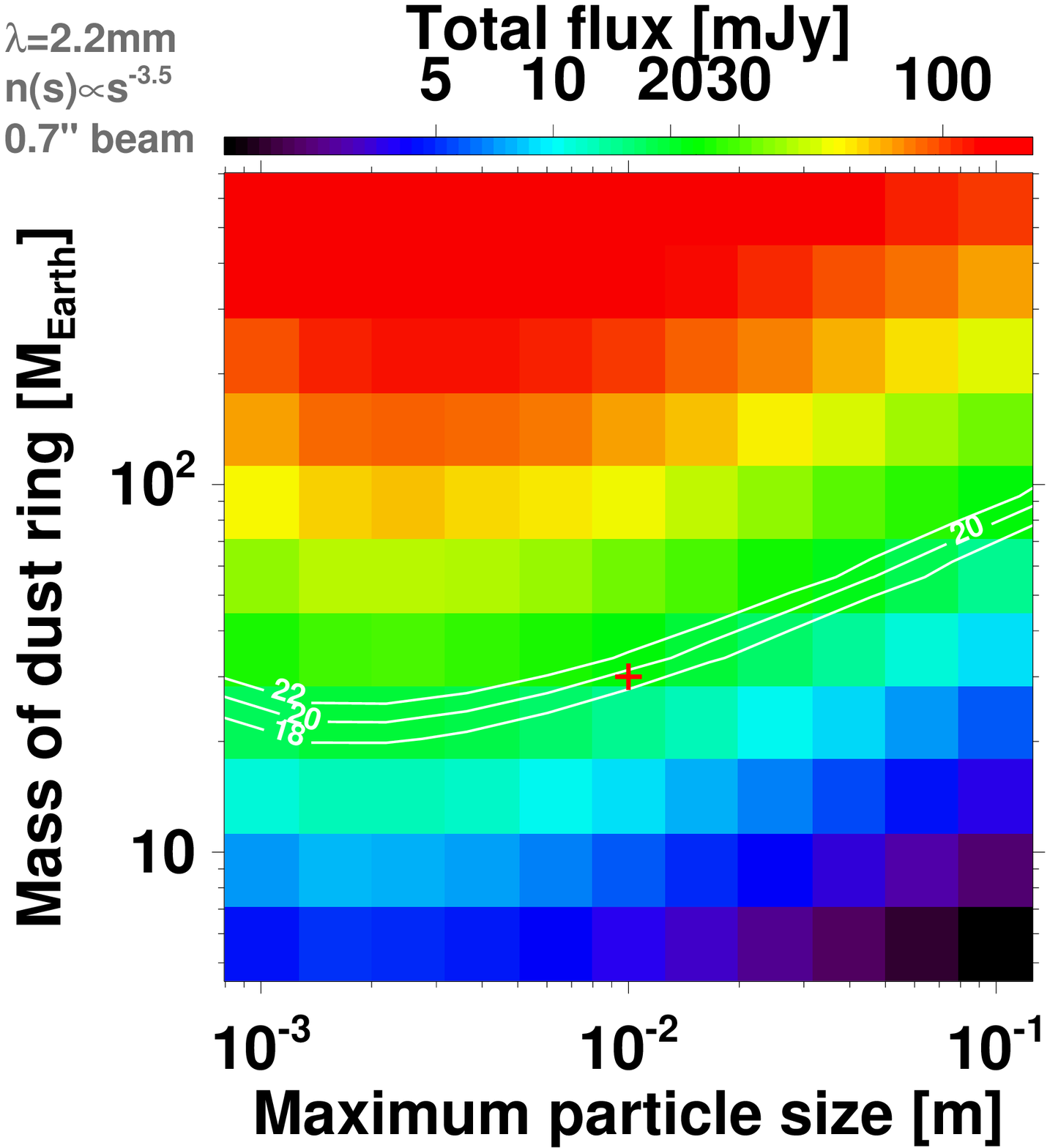}{0.45\textwidth}{(c)}
	\fig{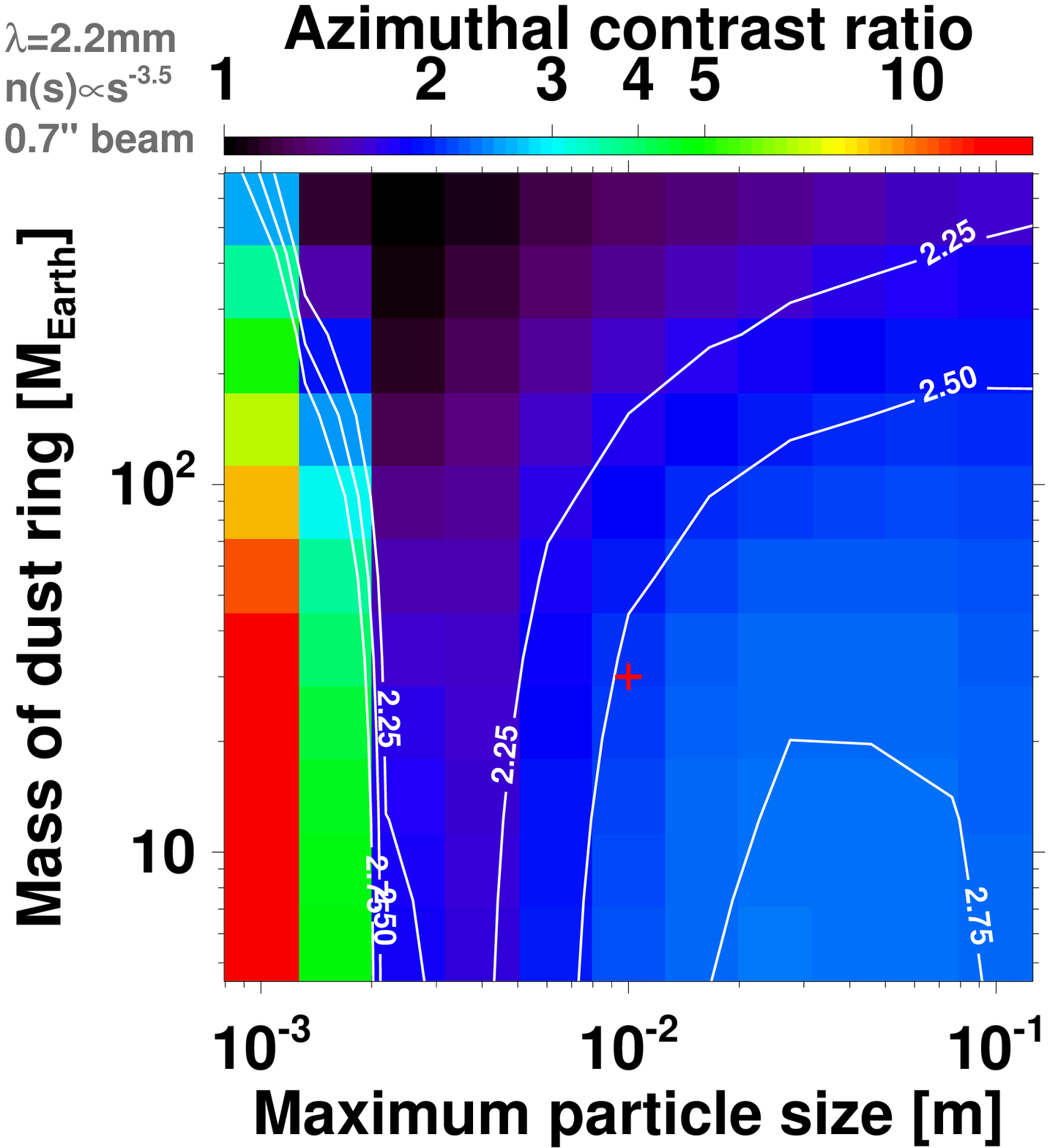}{0.45\textwidth}{(d)}
}
\caption{Total flux in the ring and azimuthal contrast ratio of the
  flux along the ring, as obtained in the synthetic maps of the dust's
  continuum emission calculated from our hydrodynamical simulations at
  370 orbits (when the vortex has started to decay).    
  The disk is assumed to be 145~pc away with an inclination of 26$^\circ$ 
  \citep{Tang2012}. Flux maps are convolved by a Gaussian beam of 0".68 FWHM to 
  compare with our observations. Results are for a size 
  distribution $n(s) \propto s^{-3.5}$, and are
  displayed against mass of dust in the ring and maximum particle
  size. The gas mass remains constant (see text).
  White contours show the approximate NOEMA values with
    10\% uncertainty, and the red cross marks our best-fit model. Panels
  (a) and (b) are at 1.12 mm, panels (c) and (d) at 2.22 mm.
  \label{Fig3}
}
\end{figure*}

We computed series of synthetic maps of the dust's continuum emission
at 1.12 and 2.22 mm in a decaying vortex (370 orbits) by varying the 
maximum particle size and the dust
mass in the ring. For each set of parameters we calculated the total
flux in the ring and the azimuthal contrast ratio of the flux along
the ring. Results are displayed in Fig.~\ref{Fig3}. In each plot,
  the white contours show the approximate NOEMA values with 10\%
  uncertainty, while the red cross marks our best-fit model, which
  corresponds to the maximum particle size (1 cm) and dust mass in the
  ring ($30M_{\oplus}$) and was used to compute the synthetic maps in
  Fig.~\ref{Fig2} to illustrate the vortex evolution.

\begin{figure*}
\centering
  \resizebox{0.99\hsize}{!}
  {
    \includegraphics{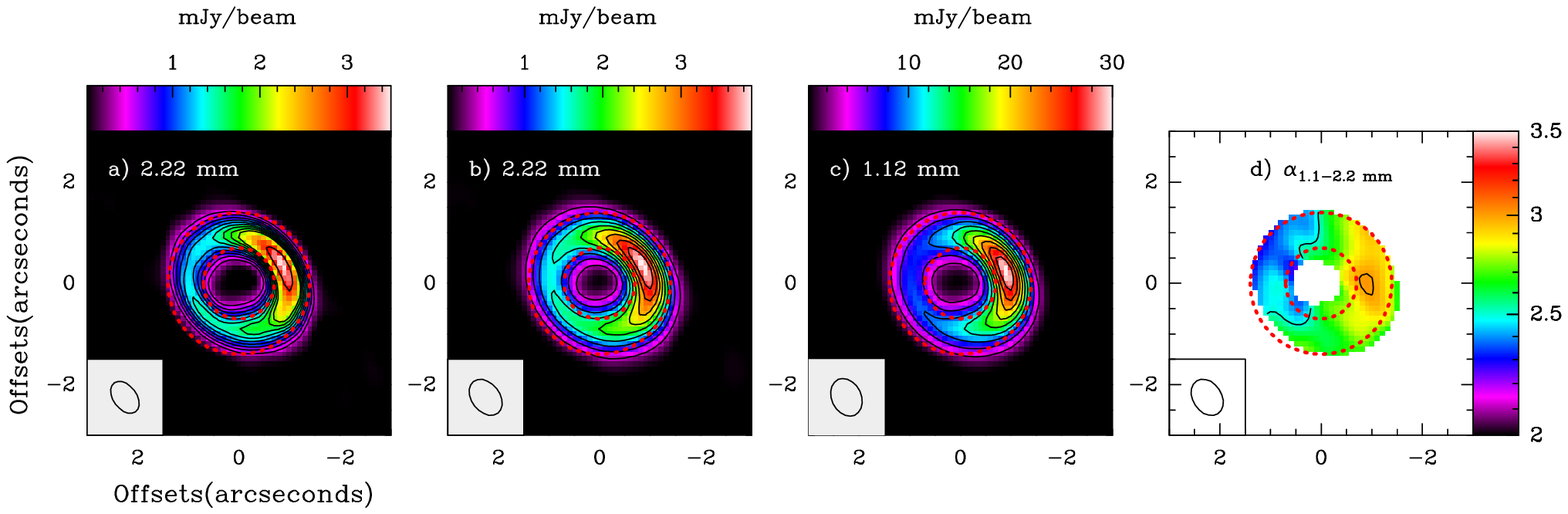}    
  }
  \caption{ Flux maps predicted by our best fit model, processed using the
    task uv$\_$fmodel of the GILDAS software to produce synthetic
    images with the same uv-coverage and weights as in our
    observations. In all panels, the beam is displayed in the
    bottom-left corner. The two red dashed circles are the same as 
    in Fig.~\ref{Fig1}.}
 \label{Fig4}
\end{figure*}

Finally, we made a detailed comparison of our best-fit model with the
observed images. Since interferometric measurements might miss flux at
the largest spatial scales, convolution of the synthetic raw flux maps
with a Gaussian beam may not properly compare with observations.
We thus processed the raw flux maps with the task uv$\_$fmodel of the
GILDAS software to produce synthetic images with the same uv-coverage
and weights as our observations.  The raw flux maps were rotated in
azimuth such that the location of the flux peak predicted at 1.12 mm
matches approximately that in the observations.  Results are shown in
Fig.~\ref{Fig4}. Our modeled flux maps are in reasonably agreement with the
NOEMA flux maps and reproduce well the intensity variations along
  the emission ring, the total flux along the ring as well as the flux
  value at the emission peak at the two wavelengths. The modeled and
observed spectral index maps (Figs.~\ref{Fig1}d and~\ref{Fig4}d)
also show overall good agreement.

\section{Concluding remarks}
\label{sec:conclusions}
Our NOEMA observations of the dust's continuum emission at 1.12 and
2.22 mm in the AB Aur disk show that the intensity variations along
the emission ring at $\approx150$ au are smaller at 2.22 mm than
at 1.12 mm. Two-fluid (gas+dust) simulations can explain this
  feature by dust trapping in a gas vortex due to a putative
2-Jupiter-mass planet companion at 96 au, provided that the
vortex has started to decay, and that solid particles are therefore
losing the azimuthal trapping effect of the vortex. Synthetic maps of
the dust's continuum emission computed from our simulations can
reproduce the total flux in, and the intensity variations along the
ring, if the trapped particles have a size distribution in $s^{-3.5}$,
a maximum size of 1 cm, and that the mass of solid particles in the
ring is $\approx 30 M_{\oplus}$. This value is consistent
  with that found by \citet[][$\sim16 M_{\oplus}$]{Tang2012} when
assuming a global dust opacity $\kappa_{{\rm 1.3mm}}$ = 0.02
cm$^{2}$ g$^{-1}$.  Future multi-wavelength, higher angular resolution
observations would be needed to confirm our findings and would allow
to accurately dissect the dust ring.

\acknowledgements { We thank the referee for a helpful and thorough
  report. We thank the Spanish MINECO for funding support from
AYA2016-75066-C2-1/2-P, AYA2016-79006-P, AYA2012-32032, FIS2012-32096 and ERC under ERC-2013-SyG,
G. A. 610256 NANOCOSMOS.

\bibliographystyle{aasjournal}

\end{document}